# Repairnator patches programs automatically

Martin Monperrus, Simon Urli, Thomas Durieux, Matias Martinez, Benoit Baudry, Lionel Seinturier

*Repairnator is a bot. It constantly monitors software bugs discovered during continuous integration of open-source software and tries to fix them automatically. If it succeeds in synthesizing a valid patch, Repairnator proposes the patch to the human developers, disguised under a fake human identity. To date, Repairnator has been able to produce patches that were accepted by the human developers and permanently merged into the code base. This is a milestone for human-competitiveness in software engineering research on automatic program repair.*

Programmers have always been unhappy that mistakes are so easy, and they need to devote much of their work to finding and repairing defects. Consequently, there has been considerable research on detecting bugs, preventing bugs, or even proving their absence under specific assumptions. Yet, regardless of how bugs are found, this vast research has assumed that a human developer would fix the bugs.

A decade ago, a group of researchers took an alternative path: regardless of how a bug is detected, they envisioned that an algorithm would analyze it to synthesize a patch. The corresponding research field is called "automated program repair."

This paradigm shift has triggered a research thrust with significant achievements [1]: Some past bugs in software repositories would have been amenable to automatic correction. In this following example, a patch fixes an if-statement with incorrect condition (a patch inserts, deletes, or modifies source code):

```
- if (x < 10)
+ if (x ≤ 10)
foo();
```

Armed with the novel mindset of repairing instead of detecting, and with opulent and cheap computing power, it is possible today to automatically explore large search spaces of possible patches.

**Program Repair Bot**

A program repair bot is an artificial agent that tries to synthesize source code patches. It analyzes bugs and produces patches, which support human developers in software maintenance. A program repair bot is potentially disruptive, because it has always been assumed that only humans can fix bugs. Repair bots can (partially) replace human developers for bug-fixing tasks.

When a bot tries to achieve a task usually done by humans, it is known as a human-competitive task [2]. The empirical evaluations of program repair research show that current program repair systems are able to synthesize patches for real bugs in real programs [3]. However, all those patches were synthesized for past bugs—those fixed by human developers in the past, usually years ago. While this indicates that automatic program repair is feasible, it is not human-competitive.

**Human-Competitiveness**

To demonstrate that program repair is human-competitive, a program repair bot has to find a good patch before a human does so. Pragmatically, that means the repair system must produce good patches in minutes, not days. Not all correct patches are good enough to be accepted by human developers, for example because they do not match the programming style. Also, the patch generated by the bot must be functionally correct compared to a patch written by a human. Note that there are patches that look correct from the test case point of view, yet that are incorrect (this is known as overfitting patches in the literature [3, 4]).
This gives us two criteria for a patch to be human-competitive: 1. the bot synthesizes the patch faster than the human developer, and 2. the human developer judges the patch to be good enough to merge into the code base.

There is one more aspect to consider. It has been shown that human engineers do not accept contributions from bots as readily as contributions from other humans, even if they are strictly identical [5]. Humans tend to be more tolerant of errors from human colleagues than from machines. This prejudice would impede our quest for a human-competitiveness proof of program repair (see **Table 1**).
To overcome this problem, we decided early in the project that our bot, which we called "Repairnator," would propose patches under a fake human identity. We created a GitHub user, called Luc Esape, presented as a software engineer in our research lab. Luc has a profile picture and looks like a junior developer, eager to make open-source contributions on GitHub. Now the human developer will not have an anti-machine reaction on receiving a patch proposal from Repairnator. This camouflage is required to test our scientific hypothesis of human-competitiveness. Luc's identity has now been disclosed.

**Automatic Repair and Continuous Integration**

Continuous integration, aka CI, is the idea that a server compiles the code and runs all tests for each commit made in the version control system of a software project (e.g. Git). In CI parlance, there is a build for each commit. A build contains the information about the source code snapshot used (e.g. a reference to a Git commit), the result of compilation and test execution (e.g. fail or success), and an execution trace log. A build fails if compilation fails or at least one test case fails. Approximately one out of four builds fail; test failure is the most common cause.

Repairnator automatically generates patches that repair build failures and shows them to human developers; a success occurs when a human developer accepts them as valid contributions to the code base. These successes would be evidence of human-competitiveness in program repair.

The setup for comparing Repairnator with human developers resembles a Turing test. When a test failure occurs in a build process, both the human developer and the bot are notified at exactly the same time. At that moment a human-bot competition begins. Repairnator is a new member of a larger class of intelligent bots for maintaining software [**6**]. Facebook has a tool called SapFix that repairs bugs found with automated testing [**7**]. DARPA's Cyber Grand Challenge (CGC) motivated the creation of very sophisticated bots for computer security [**8**]. The bots in that challenge had to be autonomous agents of two kinds: attackers, which found vulnerabilities, and defenders, which repaired vulnerabilities. Repairnator differs from the CGC defender bots two ways: Repairnator targets functional bugs, while CGC bots target security vulnerabilities; and Repairnator proposes source code patches (Java) to the developers, while CGC bots applied binary matches in machine-level code (x86).

**Repairnator in a Nutshell**

In 2017-2018, we designed, implemented, and operated Repairnator. The bot specializes on repairing build failures during continuous integration, focusing on failures due to broken tests. It constantly monitors thousands of builds on the Travis continuous integration (CI) platform and analyzes their corresponding commits. Every minute, it launches new repair attempts in order to fix bugs before the human developer. It is designed for speed because otherwise it cannot be human-competitive.

Let us now give an overview of how the Repairnator bot works.

The primary inputs of Repairnator are continuous integration builds, triggered by commits made by developers on GitHub projects—top part of the **figure**, (a) and (b). The outputs of Repairnator are two-fold: 1. it automatically produces patches for repairing builds (g), if any; 2. it collects valuable data for future program repair research (h) and (k). Repairnator monitors all continuous activity from GitHub projects (c). The CI builds (d) are given as input to a three-stage pipeline: 1. a first stage collects and analyzes CI build logs (e); 2. a second stage aims at locally reproducing the build failures that have happened on CI (f); and 3. a third stage runs different program repair prototypes coming from the latest academic research. When a patch is found, a Repairnator project member performs a quick sanity check, in order to avoid wasting valuable time of open-source developers (i). If she finds that the patch looks like a human fix (and not artificial, convoluted or surprising code), she then proposes it to the original developers of the project (j). This suggestion is done under the form of a "pull request," which is the name for a proposed code modification by an external contributor. The developers then follow their usual process of code-review and merge.
Interested readers can find more details in the publications listed in the bibliography. Repairnator focuses on repairing Java software, built with the Maven toolchain, in open-source projects hosted on GitHub, using the Travis CI platform. The fixes are generated by the Nopol [**9**], Astor [**10**], and NpeFix [**11**] algorithms. Nopol is based on dynamic analysis and code synthesis with SMT, Astor on the GenProg algorithm [**12**], and NpeFix analyzes null pointer exceptions in Java, the most common exception type.

**Expedition Achievements**

We have been operating Repairnator since January 2017, in three different phases. In January 2017, we performed a pilot experiment with an initial version of the prototype. From February through December 2017, we ran Repairnator with a fixed list of projects that we called "Expedition #1." From January through June 2018,

Repairnator monitored the Travis CI build stream in real time; the projects it analyzed were called "Expedition #2."

The main goal of the pilot experiment was to validate our design and initial implementation. We found our prototype performs approximately 30 repair attempts per day. This pilot experiment validated our core technological assumptions: A significant proportion of popular open-source projects use Travis, and the majority of them use Maven as build technology. This meant we would have a fair chance of reaching our goal of synthesizing a human-competitive patch in that context.

During Expedition #1 [13], Repairnator analyzed builds and found patches that could make the CI build successful. However, our patch analysis revealed none of those patches were human-competitive because they came too late or were of low quality. The most common cause of low quality was the limitation that a patch repairs only one buggy input.
Expedition #2 was more successful. Repairnator ran repair attempts and drafted 102 patches (available at http://l.4open.science/aa). Of those patches , Repairnator suggested 12 patches to the open-source developers, all available online. Of the 12, the developers accepted five as human-competitive. Information about those five milestone patches are shown in Table 1. and we now discuss the first of them.
The first patch merged by our program repair bot was accepted by a human developer on January 12, 2018. Here is the story: On the 12th at 12:28 p.m., a build was triggered on project aaime/geowebcache (http://l.4open.science/ab). The build failed after two minutes of execution, because two test cases were in error. Forty minutes later, at 1:08 p.m., Repairnator detected the failing build during its regular monitoring, and started to run the available program repair systems. Ten minutes later, at 1:18 p.m., it found a patch.
At 1:35 p.m., a Repairnator team member validated the proposed patch for well-formedness. At 2:10 p.m., the developer accepted the patch, and merged it with a comment: "Weird, I thought I already fixed this... maybe I did in some other place. Thanks for the patch!" That was the first patch produced by Repairnator and accepted as a valid contribution by a human developer, definitively merged into the code base. In other words, Repairnator was human-competitive for the first time.

After six more months of operation, Repairnator had processed 6,173 failures and drafted 12 patches, of which five accepted and merged by human developers. We reproduce verbatim below the patch for the software application Ditto. This automatic patch adds a precondition to guard a statement responsible for a crash. The condition of the precondition is synthesized with the Nopol program repair algorithm [9]

```
    LogUtil.enhanceLogWithCustomField(log,…
-   stopMessageMappingProcessorActor();
+   if (this.isConsuming()) {
+     stopMessageMappingProcessorActor();
+   }
```

There were two reasons that the other seven proposed patches were not accepted. There was no answer from the owner, probably because the project did not accept external contributions; or the owner asks for more changes in the code, which is out-of-reach of Repairnator's capabilities.

Overall, the Repairnator project has fulfilled its mission. It demonstrated program repair can be human-competitive—patches of good quality delivered before humans themselves found patches.

**The Future**

This early experiment is a proof-of-concept that human-competitive patches can be generated automatically. We believe automatic program repair can become common and generate large cost savings in software development.

Our early experiment produced five accepted patches for 6,173 failures. Obviously, this is not serious relief to human developers making patches. But given there are on the order 12 million bugs detected globally at an average cost of $400 per fix, the potential is large and even the acceptance rate we observed in our experiment represents a savings of $3.8 million globally.

With the current growth of active research in that area, bots will get ever better at bug fixing and the success rate of automatic repairs will steadily raise over the next few years. Funding agencies and calls, such as DARPA's MUSE, are providing opportunities for more research in this area.

Of course, there is a limit. Some bugs cannot be fixed without deep algorithmic or architectural understanding. Such bugs will remain out of reach for automatic repair for the foreseeable future.

As they become more common and more successful, repair bots will work side by side. Repair bots will fix the easiest bugs, which are larger in number, leaving the toughest repairs for humans. In between, we expect a gray area of bugs where bots will propose a partial fixes and humans will complete the jobs. This can work the other way around too: Humans could make initial fixes and repair bots would examine and validate them [14]. Automatic bug fixing may have an impact on the psychology of programmers. Developers could become sloppier because they would count on bots to catch stupid mistakes. We take a more optimistic view: Repairs bots will remove a tedious part of software development. Given the current shortage of software developers, which is going to last, making software development more attractive, more creative, and more fun again would be an excellent byproduct of automatic bug fixing.

In addition to showing program repair is human competitive, the Repairnator project has provided a wealth of information about bugs and continuous integration, which is made available for the research community.

**Acknowledgment**

This work was partially supported by the Wallenberg Artificial Intelligence, Autonomous Systems and Software Program (WASP) funded by Knut and Alice Wallenberg Foundation and by the InriaHub program.

**Authors**

Martin Monperrus is Professor of Software Technology at KTH Royal Institute of Technology. He was previously associate professor at the University of Lille and adjunct researcher at Inria. He received a Ph.D. from the University of Rennes, and a master's degree from the Compiègne University of Technology. His research lies in the field of software engineering with a current focus on automatic program repair, program hardening, and chaos engineering.

Simon Urli received a Ph.D. in software engineering in 2015 from the University of Nice-Sophia Antipolis while he was working on composition of software product lines. He then worked as a postdoc at University of Lille between 2016 and 2018 on the topic of automatic software repair and was deeply involved in the creation of Repairnator. He is now a software engineer at XWiki SAS since 2018.

Thomas Durieux is a postdoc at INESC-ID (Portugal) since 2019, and he received his Ph.D. in computer science from University of Lille (France). His current research interest is about the automation of software engineering tasks such as patch generation and fault localization with dynamic and static techniques.

Matias Martinez is an assistant professor at the Université Polytechnique Hauts-de-France (France). He received his diploma in computer science from UNICEN (Argentina) and his Ph.D. from University of Lille (France). His research interests are focused on software evolution, automated software repair, software testing and mobile applications.

Benoit Baudry is a Professor in Software Technology at the KTH Royal Institute of Technology, where he leads the CASTOR software research center. He received his Ph.D. in 2003 from the University of Rennes and was a research scientist at INRIA from 2004 to 2017. His research interests include dynamic code analysis, automated software engineering and software diversification.

Lionel Seinturier is Professor of Computer Science at the University of the University of Lille, France, since 2006. He is heading the Computer Science department at the University of Lille, and the Inria Spirals project-team. His research interests include distributed systems and self-adaptive software systems. From 1999 to 2006, he was Associate Professor at Sorbonne University, France. He earned his Ph.D. in 1997 from CNAM, Paris. He is the author of two books and more than 80 publications in international journals and conferences. He participated to the development of four major software platforms. He was member of program committees of more than 60 international conferences. He is or has been involved in 24 funded research grants (EU, industry, national agencies). Since 2005, he has directed and is directing 19 Ph.D. theses.

## Figures

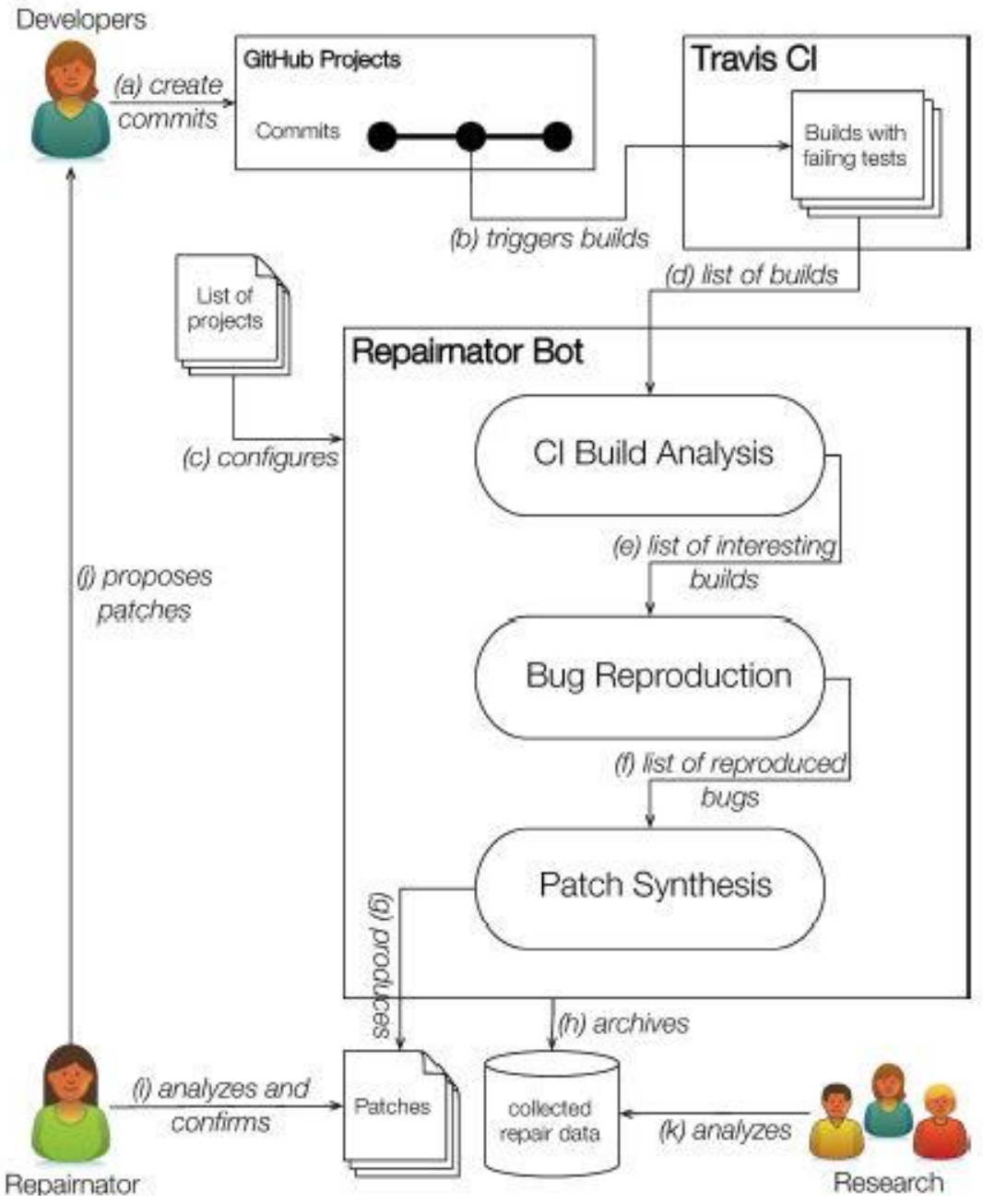

Figure 1. The Repairnator workflow.

## Tables

| Date | Contribution | Developer comment |
|---|---|---|
| January 12, 2018 | aaime/geowebcache/pull/1 | "Thanks for the patch!" |
| March 23, 2018 | parkito/BasicDataCodeU[...]/pull/3 | "merged commit 140a3e3 into parkito:develop" |
| April 5, 2018 | dkarv/jdcallgraph/pull/2 | "Thanks!" |
| May 3, 2018 | eclipse/ditto/pull/151 | "Cool, thanks for going through the Eclipse process and for the fix." |
| June 25, 2018 | donnelldebnam/CodeU[...]/pull/151 | "Thanks!!" |

Table 1. Five Repairnator patches accepted by the developer community.